\documentclass[conference,a4paper]{IEEEtran}

\usepackage{url}
\usepackage{cite}
\usepackage{amsthm}
\usepackage{amsmath}
\usepackage{amssymb}
\usepackage{amsfonts}
\usepackage{graphicx}
\usepackage{subfig}

\newtheorem{definition}{Definition}

\newtheorem{remark}{Remark}
\newtheorem{lemma}{Lemma}
\newtheorem{theorem}{Theorem}

\usepackage{algorithm,algorithmicx,algpseudocode}
\algrenewcommand{\algorithmiccomment}[1]{\hfill$\blacktriangleright$ #1}

\usepackage{times} 

\usepackage[right=0.5in,left=0.5in,top=.7in,bottom=1in]{geometry}

\newcommand\co[1]{}

\newcommand\FF{\mathbb{F}}
\newcommand\bv[1]{\mathbf{#1}}

\mathchardef\d="2D

\newcommand\bra[1]{\left\{#1\right\}}
\newcommand\inb[1]{\left(#1\right)}

\newcommand\hw[1]{\operatorname{w}\!\left(#1\right)}

\newcommand\brank[1]{\operatorname{rank}_2\left(#1\right)}

\begin{document}
\title{Locally Repairable Codes with\\ Multiple Repair Alternatives}

\author{
  \IEEEauthorblockN{Lluis Pamies-Juarez, Henk D.L. Hollmann and Fr\'ed\'erique Oggier}
  \IEEEauthorblockA{School of Physical and Mathematical Sciences\\
    Nanyang Technological University\\
    Singapore\\
    Email: \{lpjuarez,henk.hollmann,frederique\}@ntu.edu.sg} 
}

\maketitle

\begin{abstract} Distributed storage systems need to store data redundantly
in order to provide some fault-tolerance and guarantee system reliability.
Different coding techniques have been proposed to provide the required
redundancy more efficiently than traditional replication schemes.  However,
compared to replication, coding techniques are less efficient for repairing
lost redundancy, as they require retrieval of larger amounts of data from
larger subsets of storage nodes. To mitigate these problems, several recent
works have presented locally repairable codes designed to minimize the repair
traffic and the number of nodes involved per repair. Unfortunately, existing
methods often lead to codes where there is only one subset of nodes able to
repair a piece of lost data, limiting the local repairability to the availability of the nodes in this subset.

In this paper, we present a new family of locally repairable codes that
allows different trade-offs between the number of contacted nodes per repair,
and the number of different subsets of nodes that enable this repair.  We
show that slightly increasing the number of contacted nodes per repair allows
to have repair alternatives, which in turn increases the probability of being
able to perform efficient repairs.

Finally, we present $pg$-BLRC, an explicit construction of locally repairable
codes with multiple repair alternatives, constructed from partial geometries,
in particular from Generalized Quadrangles. We show how these codes can
achieve practical lengths and high rates, while requiring a small
number of nodes per repair, and providing multiple repair alternatives. 
\end{abstract}

\section{Introduction}

In recent years, distributed storage systems as used in large data centers
have started to incorporate coding techniques to redundantly store data
across different storage nodes. For example, Facebook reported that it is 
archiving old data using a classic Reed-Solomon code implemented on top of
the Hadoop Distributed File System (HDFS)~\cite{warehouseface,diskreduce2},
and Microsoft uses a Pyramid Code as the main storage primitive of its
Azure storage service~\cite{azureec}. The use of coding mechanisms in these
distributed storage systems provides significantly higher fault-tolerance
values and lower storage overheads than simpler replication
schemes~\cite{highavail}. For example, in systems like HDFS or Azure, coding techniques allow to store data with a footprint of 1.3--1.5 times
the size of the original object, which represents a footprint reduction of
50\% as compared to the \emph{de-facto} standard 3-replica scheme.

The main problem of using traditional coding techniques in distributed
storage systems is that repairing lost encoded data requires the retrieval of
large amounts of data from large subsets of nodes, which entails an important
network traffic and lots of input/output (I/O) operations.  In today's large
distributed storage systems where node failures are the norm rather than the
exception, minimizing the communication costs due to data repairs has
therefore become an important problem. Regenerating Codes~\cite{Dimakis} were
the first families of a new wave of codes especially designed to minimize repair
costs in distributed storage systems. Regenerating Codes address the repair
problem by describing an optimal trade-off between the storage overhead of
the code and its repair communication costs. However, repair processes in
Regenerating Codes require contacting a large subset of nodes, which
complicates the design of the storage system and increases the number of
required I/O operations. 

A different line of new codes, called locally repairable codes (LRC), also
addresses the repair problem, but focusing on reducing the number of nodes
contacted during repair~\cite{papailiopoulos2012locally,
kamath2012codes,OD,ODITW,mdslocal, rawat2012locality}, while still
guaranteeing a low repair traffic. However, the main problem with existing
locally repairable codes is that although they reduce the size of the subset
of contacted nodes, they suffer from the drawback that only a single subset
of nodes enables the repair of a specific piece of redundant data.  If a
single node from this repair subset is not available, data cannot be
repaired ``locally'', increasing the cost of the repair. Alternatively, some
earlier codes like Pyramid codes~\cite{huang2007pyramid}, or
Hierarchical codes~\cite{hierarchicalcodes}, provide different subsets of
nodes that enable the repair of each piece of redundant data. However, these
different repair subsets do not have all the same size: during normal
operations, lost data can be repaired using the smallest subset, however,
when the number of unavailable nodes grows, repairs require the use of larger
subsets, thus increasing the repair cost.

In order to maximize the reliability of storage systems it is therefore
desirable to obtain codes where lost data can be repaired by contacting a
small number of nodes $r$, where this number can be as small as $r=2$. In
addition, since the repair process might not be able to contact some of these
$r$ nodes during repair (e.g., due to temporary node unavailabilities, or
even a correlated failure of multiple nodes), it is also desirable to have
$a>1$ different alternative $r$-subsets of nodes enabling the repair of any
lost data.  Unfortunately, although some locally repairable
codes~\cite{OD,ODITW} present constructions where $a>1$, to the best of our
knowledge there is no publication focusing on the analysis and design of
codes that allow a {\em trade-off} between the values of $a$ and $r$.

\vspace{.75em}
\emph{Contributions:}

In this paper, we present a new framework to facilitate the analysis and
design of locally repairable codes with different trade-offs between their
repair locality and their number of repair alternatives per failure. This
framework allows us to define the ``local repair tolerance'' as a new metric
that measures the maximum number of nodes that can be unavailable in the
system without compromising the ability to locally repair all the stored
data. Furthermore, we present $pg$-BLRC codes, an explicit construction of
locally repairable codes with multiple repair alternatives, constructed from
partial geometries.  We provide an upper and lower bound for the rate of such
codes, which is attained by the class of Generalized Quadrangles.  Design of
efficient $pg$-BLRC codes involves a trade-off among three of its desirable
features: (i) small repair locality, (ii) large number of repair
alternatives, and (iii) low storage footprint. Throughout numerical evaluations
we identify that optimizing individually each of these properties leads to a
bad performance of the other two. Requiring multiple repair alternatives
thus introduces a new point of view in the study of the optimal trade-offs
between the rate and the repair locality of storage codes.


\section{Linear Codes for Distributed Storage}
\label{s:lin}

Let $q$ be a prime power. A $q$-ary linear code $C$ of length $n$ and rank $k$ is a $k$-dimensional linear subspace of
the vector space $\FF_q^n$, where $\FF_q$ is the finite field
with $q$ elements and $n\geq k$. As a linear subspace, the code $C$ can
be defined by a $k\times n$ full-rank generator matrix $G$ as $$C=\bra{\bv oG:
\bv o\in\FF_q^k},$$ where the vectors $\bv c\in C$ are called the
\emph{codewords} of $C$. The code $C$ can alternatively be defined using an
$(n-k)\times n$ full-rank parity check matrix $H$ such that $GH^\top=0$, as
$$C=\bra{\bv c\in\FF_q^n:H\bv c^\top=0}.$$

In distributed storage systems, a data object of $k$ symbols is represented as
a vector $\bv o\in\FF_q^k$, containing $k\lceil\log_2 q\rceil$ bits. To
redundantly store this object across different storage nodes, the system first
obtains the codeword $\bv c= \bv o G\in\FF_q^n$, and then it
stores the $n$ symbols of $\bv c$ on $n$ different storage nodes. Since
$n\geq k$, each stored object requires a disk capacity larger than its
original size. The rate $R=k/n$ of the code represents the
proportion of storage capacity that is used to store non-redundant symbols.
The closer $R$ is to one the more storage-efficient the code is. The
efficiency of codes is also measured in terms of the storage footprint,
$n/k$, which is the capacity used to store each data object compared to its
original size.

After storing the codeword $\bv c$ the system can reconstruct the original
object $\bv o$ by gathering some $k$ symbols out of the $n$ stored ones. For
that, let $\mathcal I$, $|\mathcal I|=k$, be a set containing the indexes of
these $k$ symbols. Then the object $\bv o$ can be reconstructed by solving
the system $\bv o= \bv c_\mathcal I G_\mathcal I^{-1}$, where $\bv c_\mathcal
I$ is the vector composed of the elements of $\bv c$ that are indexed by the
members of $\mathcal I$, and similarly, $G_\mathcal I$ is the submatrix of
$G$ composed of the columns of $G$ that are also indexed by the members of
$\mathcal I$. It is important to note that the previous system can only be
solved if the matrix $G_\mathcal I$ is invertible. When this matrix is
invertible for any $k$-subset $\mathcal I$ we say that the code is a maximum
distance separable code (or MDS code).

\section{Data Repairability of Linear Codes}
\label{s:rep}

As pointed out in the introduction, providing efficient mechanisms to
repair lost encoded data is an important problem in distributed storage
systems. In this section we introduce an analytic framework to evaluate the
repairability properties of different linear codes, focusing on codes
providing local repairs.

Let the code $C^\bot$, the dual of $C$, be the code generated by
the parity check matrix $H$. Then, by definition all codewords $\bv v\in
C^\bot$ are parity check vectors of $C$, which means that $\bv v\bv c^\top=0$
for any $\bv c\in C$. This parity check property can be used
to construct repair mechanisms able to repair lost symbols in the codewords of $C$. For
that, let $\bv v \in C^\bot$ be a parity check vector with a nonzero $i$th
symbol, that is $\bv v(i)\ne 0$. Then, repairing the $i$th symbol of
$\bv c$, $\bv c(i)$, consists of solving the equation $\bv v\bv c^\top=0$, or
equivalently, solving
$$
\bv v(1)\bv c(1)+\bv v(2)\bv c(2)+\dots+\bv v(n)\bv c(n) = 0.
$$
This equation has then as many unknowns as number of nonzero symbols in $\bv
v$, or $\hw{\bv v}$ unknowns, where $\operatorname{w}$ is the Hamming weight
function. Then, repairing a missing symbol of $\bv c$ requires to retrieve $\hw{\bv
v}-1$ other symbols and solve the equation for $\bv c(i)$. However,
retrieving $\hw{\bv v}-1$ symbols over a communication network might entail a
significant overhead in terms of network traffic. Consequently, to minimize
this traffic it is important to design codes $C$ guaranteeing that for every
$i\in[n]=\{1,2,\dots, n\}$ there exists at least one vector $\bv v\in C^\bot$ with $\bv
v(i)\ne0$, and small Hamming weight $\hw{\bv v}$.

To analyze the repair efficiency of codes we can enumerate all the possible
ways in which the $i$th symbol of a codeword $\bv c\in C$ can be repaired.
To that end, we define $\Omega(i)$ as the set containing all the parity check
vectors repairing this symbol:
$$
\Omega(i) = \bra{\bv v\in C^\bot:\bv v(i)\ne 0}.
$$
Then, for each $i\in[n]$, we can evaluate the
cost of repairing the $i$th symbol by analyzing the Hamming weight of all
vectors $\bv v\in\Omega(i)$.  The \emph{repair degree} is a metric that
describes the number of symbols that need to be retrieved per repair:

\begin{definition}[Repair Degree]
We define the repair degree for the $i$th codeword symbol as $r(i) =
\min\bra{\hw{\bv v}-1:\bv v\in\Omega(i)}$, and the overall repair degree $r$
of a linear code is its maximum repair degree: $r=\max\bra{r(i)}_{i=1}^n$.
\label{d:r}
\end{definition}

In classic MDS codes such as Reed-Solomon codes~\cite{reedsolomon}, or in
Regenerating Codes~\cite{Dimakis}, we have that the repair degree is at least
equal to the rank of the code, $r\geq k$. For Reed-Solomon codes it means
that repairing a single failure requires to transfer an amount of information
equal to the size of the original object, or $k\lceil\log_2 q\rceil$ bits.
For Regenerating Codes, although they still have to contact at least $k$
nodes per repair, the overall amount of data transferred per repair can be
slightly reduced below the size of the original object. However, instead of
aiming at reducing the repair traffic, in this paper we are interested in
codes able to improve the repair performance by reducing the number of nodes
that need to be contacted per repair below $k$, which in turn leads to reduce
the network traffic per repair as well. We will refer to those codes with an
overall repair degree much smaller than its rank ($r\ll k$) as \emph{locally
repairable codes}, or LRC.

\section{Codes with Multiple Repair Alternatives}
\label{s:cod}

Even though locally repairable codes significantly reduce the number of nodes
that need to be contacted during repairs, it is often also desirable to
guarantee the existence of multiple subsets of this kind. This is especially
important in storage systems where some of the nodes to be contacted might be
unavailable, either because they are temporarily busy, or because there was a
correlated failure affecting several nodes. In this section we measure the
multiple repair alternatives of LRC codes and their ability to perform local
repairs in the presence of node unavailabilities.

Let $\Omega_r(i)$ be the subset of $\Omega(i)$ containing the vectors that
allow to repair $\bv c(i)$, $\bv c\in C$, with a repair degree at most $r$,
i.e., $\Omega_r(i)=\bra{\bv v\in\Omega(i):\hw{\bv v}\leq r+1}$. Each of these
vectors represents then a possible alternative to repair the $i$th symbol of
any codeword $\bv c\in C$. 

\begin{definition}[Repair Alternativity]
The repair alternativity of the $i$th codeword symbol is the number of
distinct subsets of nodes with at most $r$ nodes, which contain enough
information to repair the $i$th symbol. The repair alternativity of $i$ is then
$a(i)=|\Omega_r(i)|$, and the code's overall repair alternativity
is $a=\min\bra{a(i)}_{i=1}^n$.
\label{d:a}
\end{definition}

Non-locally repairable MDS codes such as Reed-Solomon codes and Regenerating
Codes have a large repair alternativity of $a=\binom{n}{r}$, which guarantees
that all stored symbols can be repaired even when a large portion of storage
nodes is unavailable. In fact, all symbols can be repaired as long as the
stored information is available (that is, if $k$ symbols are available).  On
the other hand, most of the existing locally repairable
codes~\cite{papailiopoulos2012locally,kamath2012codes} have a local repair
alternativity of $a=1$. In this case, if any of the $r$ nodes involved in the
repair is temporary unavailable, the code cannot use the local repair
mechanisms, requiring then more expensive repair solutions.  To the best of
our knowledge SRC~\cite{OD,ODITW} are the only LRC codes with a repair
alternativity larger than one, $a>1$. Unfortunately, in SRC codes the value
of $a$ depends on the code construction, which does not allow to obtain codes
with arbitrary $a$ values. Moreover, obtaining SRC codes with large $a$ leads
to unpractical codes with low rate $R$. 

In this paper, we focus on the design of locally repairable codes with
arbitrary repair alternativity and practical rates. Having codes with
multiple repair alternatives increases the probability to be able to locally
repair lost data when some nodes are unavailable. To maximize the local
repair probability, it is therefore important to guarantee that the number of
common nodes in different repair alternatives for a given symbol is as small
as possible.  For example, in order to maximize the number of repair
alternativities of the $i$th symbol, $a(i)$, we have to minimize
$|\operatorname{supp}(\bv v)\cap\operatorname{supp}(\bv u)|$, for any
distinct pair of vectors $\bv v,\bv u\in\Omega_r(i)$, where
$\operatorname{supp}(\bv v)$ is the support of $\bv v$, which is the set
containing the indices of the non-zero positions of $\bv v$. Using this
concept, we can formally define the local repair tolerance of a LRC code
as follows:

\begin{definition}The local repair tolerance of the $i$th symbol,
$\delta(i)$, is the size of the smallest set of coordinates different than $i$
that intersects with the support of all codewords in $\Omega_r(i)$:
$$
\delta(i)\!=\!\min\bra{|\mathcal I|:\mathcal I\!\subset\![n]\!\setminus\!\{i\},~\mathcal
I\cap\operatorname{supp}(\bv v)\ne\emptyset,~\forall~\bv
v\!\in\Omega_r(i)}\!\!.
$$
The overall local repair tolerance of the code is \linebreak
$\delta = \min\bra{\delta(i)}_{i=1}^n$.
\label{d:tol}
\end{definition}

This means that the $i$th symbol can be locally repaired when at most
$\delta(i)$ nodes are unavailable, and any symbol is locally repairable when
at most $\delta$ nodes are unavailable.  In the design of locally repairable
codes we will then aim at maximizing the value of $\delta$. However, some code
designs might have unbalanced constructions where $\delta(i)\ll\delta(j)$ for
different symbols $i,j\in[n]$, where different symbols have different
probability to be repaired locally.  To avoid working with this unbalanced
code constructions, in this paper we only focus on balanced locally
repairable codes:

\begin{definition}[Balanced Codes]
When $\delta(i)=\delta(j)$ for all \linebreak $i,j\in[n]$, $i\ne j$, we say
that the code is balanced.
\end{definition}

We will refer to these balanced locally repairable codes as BLRC codes, and
we will use the notation $(n,k,r,a,\delta)$-code to refer to a code of length
$n$ and rank $k$, where each symbol of the codeword can be repaired from at
most $r$ other symbols, having at least $a$ different sets of symbols that
guarantee such a repair, and being able to locally repair each symbol if
there are at most $\delta$ unavailable symbols. In the next section we will
present a simple code construction of BLRC code with arbitrary $a$ and $r$
values. 

\section{Balanced Locally Repairable Codes \\from Partial Geometries}
\label{s:blrc}

We present a way to generate explicit BLRC constructions from partial
geometries and analytically evaluate
the new codes in terms of repair tolerance $\delta$ and code rate $R$.
We first provide a brief description of partial geometries.

\subsection{Partial Geometries}

A partial geometry $pg(s,t,\alpha)$ is an incidence structure between a set of
points $\mathcal P$ and a set of lines $\mathcal B$ such that:
\begin{enumerate}
\item Each point $P\in\mathcal P$ is incident with $t+1$ lines ($t\geq1$).
\item Each line $B\in\mathcal B$ is incident with $s+1$ points ($s\geq1$).
\item Any two lines have at most one point in common.
\item If a point $P$ and a line $B$ are not incident, there are exactly
$\alpha$ ($\alpha\geq1$) pairs $(Q,M)\in\mathcal P\!\times\!\mathcal B$, such
that $P$ is incident with $M$ and $Q$ is incident with $B$.
\end{enumerate}

It follows that $1\leq\alpha\leq\min\bra{t+1,s+1}$,
and by definition of partial geometries the cardinalities of the point and
line sets must satisfy:
\[
|\mathcal P| = \frac{(s+1)(st+\alpha)}{\alpha}
\text{ and }
|\mathcal B| = \frac{(t+1)(st+\alpha)}{\alpha}.
\]

The dual of a partial geometry, which is the incidence structure that arises from
interchanging the set of points $\mathcal P$ with the set of lines $\mathcal
B$, is also a partial geometry with parameters $pg(s'=t,t'=s,\alpha)$. Finally,
according to the values of the parameters $s$, $t$ and $\alpha$, partial
geometries can be divided into four classes:
\begin{enumerate}
\item When $\alpha=s+1$, or dually $\alpha=t+1$, the partial geometry is a
Steiner 2-design.
\item When $\alpha=s$, or dually $\alpha=t$, the partial geometry is called a
\emph{net} or a \emph{transversal design}.
\item When $\alpha=1$, the partial geometry is called a \emph{generalized
quadrangle}.
\item For $1<\alpha<\min\bra{s,t}$ the partial geometry is \emph{proper}.
\end{enumerate}
As we will show in Section~\ref{l:Ropt}, generalized quadrangles are of special
interest to design optimal codes in terms of rate.

\subsection{Codes from Partial Geometries}
\label{s:cons}

Partial geometries and other incidence structures have been widely studied
for the construction of LPDC codes~\cite{li2008regular,johnson2004codes}. The
incidence matrix of partial geometries can be used as a simple mechanism to
obtain sparse parity-check matrices with low rank over $\FF_2$, which makes
them particularly suitable to construct high rate LPDC codes with efficient
iterative decoders. The similarity of the requirements of these LPDC codes
with those of BLRC codes makes incidence matrices of partial geometries a
promising source of BLRC designs. We will use some of the results of
Johnson and Weller~\cite{johnson2004codes} to evaluate the local
repairability of such codes. Although in the previous sections we considered
generic $q$-ary codes, in this section we limit our code designs to binary
codes, i.e., $q=2$. As we will show, this limitation does not affect the rate
of the obtained code.
 
Let us define the incidence matrix of a partial geometry $pg(s,t,\alpha)$ as
a $|\mathcal B|\times|\mathcal P|$ matrix $N=(n_{ij})$, where $n_{ij}=1$ or
$0$ according as whether the $i$th line is incident with the $j$th point or
not.  Constructing a linear code $C$ from a partial geometry consists then of
building an $m\times n$ parity check matrix $H$ containing $m$ linear
independent rows of $N$, where $m=\brank{N}$. Then we can obtain the
generator matrix $G$ of the code $C$ by solving the equation $GH^\top\!=0$. If
$H$ can be expressed as $H=[I_{n-k},Q]$, the
generator matrix is then defined as $G=[-Q^\top|I_k]$.
Note that this requires that $Q$ cannot contain all zero rows.
And since the generator matrix $G$ contains an identity matrix, the code
$C$ is systematic. We will denote a code $C$ constructed from a
partial geometry $pg(s,t,\alpha)$, as a $pg$-BLRC code.

Besides the formal definition of $pg$-BLRC codes, we can also state the
following lemma regarding the repair degree and the repair alternativity of
such codes.

\begin{lemma}
The repair degree of a $pg$-BLRC code $C$ and its repair alternativity
satisfies $r\leq s$, and $a\geq t+1$.
\label{l:l}
\end{lemma}

\begin{IEEEproof}
From the properties of partial geometries we have that every point
$P\in\mathcal P$ is incident with $t+1$ lines, and each of these lines is at
the same time incident with $s+1$ points.  It means that for each $i\in[n]$
there are $t+1$ rows of the incidence matrix $N$, namely $\bv v_0\dots\bv
v_t$, such that $\bv v_j(i)=1$ and $\hw{\bv v_j}=s+1$ for all $j\in[n]$.
Then, by definition of $\Omega_r(i)$ we have that $\bv v_j\in \Omega_r(i)$,
for all $j=0,\dots,t$, and hence $|\Omega_r(i)|\geq t+1$. Then, from
Definition~\ref{d:a} we get that $a\geq t+1$, and similarly, from
Definition~\ref{d:r} we get that $r(i)\leq s$ for all $i\in[n]$ and thus
$r\leq s$.
\end{IEEEproof}

For the rest of the paper we will call ($r,a$) $pg$-BLRC codes those
codes constructed from partial geometries $pg(s,t,\alpha)$, where $s>1$. When
$s=1$ any lost data can be repaired by contacting a single node in the system,
which corresponds to a simple data replication scheme. The condition $s>1$
allows us to exclude replication schemes from the definition of our BLRC codes.

\begin{lemma}
A $pg$-BLRC code with $s>1$ has a per-symbol repair tolerance bounded by
$\delta(i)\geq t+1$, for all $i\in[n]$, and an overall repair tolerance bounded
by $\delta\geq t+1$. 
\label{l:tol}
\end{lemma}

\begin{IEEEproof}
Let $\mathcal N(i)\subseteq\Omega_r(i)$ be the set containing all the rows
$\bv v$ of the incidence matrix $N$ satisfying that $\bv v(i)=1$, for all
$i\in[n]$.  On the one hand, let us first assume that $\mathcal
N(i)=\Omega_r(i)$. In this case, from property 3) of partial geometries we
have that for any two $\bv v_1,\bv v_2\in \mathcal N(i)$, $\bv v_1\ne \bv v_2$,
the support intersection $\operatorname{supp}(\bv v_1)\cap \operatorname{supp}(\bv v_2)$ is
either the empty set or the set $\{i\}$.
Then, since $|\mathcal N(i)|=t+1$, the minimum subset $\mathcal I\subseteq
[n]\setminus\{i\}$ that intersects the support of all vectors in $\mathcal N(i)$ satisfies $|\mathcal I|=t+1$, and then from Definition~\ref{d:tol} it
follows that $\delta(i)=a(i)=t+1$ and $\delta=a=t+1$. On the other hand, if
we assume $\mathcal N(i)\subset\Omega_r(i)$, then the extra vectors $\bv
w\in\Omega_r(i)\setminus\mathcal N(i)$ might only contribute to increase the
repair tolerance of the symbol $i$, and hence $\delta(i)\geq t+1$.
\end{IEEEproof}

From the last two lemmas we know that an ($r,a$) $pg$-BLRC code $C$
constructed from a $pg(s,t,\alpha)$ partial geometry guarantees that $r\leq
s$, $a\geq t+1$ and $\delta\geq t+1$. Not achieving these three bounds with
equality implies that there exists some codewords in the dual code $C^\bot$
with a Hamming weight smaller than $s+1$. Due to the difficulty of finding
such a type of codewords, we will refer to an ($r,a$) $pg$-BLRC code as code
with a \emph{designed} repair degree $r$, and a \emph{designed} repair
alternativity $a$. This gives to the designer of the storage system the
guarantee to be able to repair \emph{all} missing symbols by contacting $r$
other nodes, having $a$ alternative $r$-subsets that enable this repair (the
lines of the geometry), and the guarantee that the system can repair any
failure when at most $\delta=a$ nodes are temporary unavailable.

\subsection{Rate Bounds of $pg$-BLRC Codes}
\label{l:Ropt}

In the previous sections we have presented the construction of $pg$-BLRC codes
and the properties that allow to measure the repair performance of these codes.
However, besides offering efficient repair mechanisms, codes used in
distributed storage systems also need to guarantee low storage overheads (or
equivalently high code rates). In this section we provide an upper and lower bound for
the rate $R$ of ($r,a$) $pg$-BLRC codes.

\begin{theorem}
The rate $R$ of an ($r,a$) $pg$-BLRC code $C$ is lower bounded by
\vspace{-3mm}
$$
R\geq \frac{r^2}{(a+r-1)(r+1) },
$$
and when $r+a-1$ is even, the rate is upper bounded by
$$
R\leq\frac{a (r^2-r+1) -(r-1)^2}{(a + r-1) (r(a-1)+1)}.
$$
\label{t:R}
\end{theorem}
\vspace{-4mm}
\begin{IEEEproof}
The rate of an ($r,\!a$) $pg$-BLRC code is 
\begin{equation}
R=\frac{k}{n}=\frac{n-\brank{N}}{n},
\label{e:R}
\end{equation}
where $N$ is the incidence matrix of the partial geometry. From
Johnson~et~al.~\cite{johnson2004codes} we have that the $\brank{N}$ of such
codes is upper bounded by $\brank{H} \leq \vartheta+1$, and when
$s+t+1-\alpha$ is even, then is is lower bounded by $\vartheta \leq
\brank{H}$, where
\[
\vartheta = \frac{st(s+1)(t+1)}{\alpha(t+s+1-\alpha)}.
\]
{\em Lower Bound:} The minimum possible rank is achieved when
$\brank{N}=\vartheta+1$.  Substituting in (\ref{e:R}) $\brank{N}$ by
$\vartheta+1$, and $n$ by $|\mathcal P|$ we get:
$$
R\geq \frac{-s \alpha  + s(s-1)~~~~~~~~~~~}{ -(s+1)\alpha + s(s+2) + t(s+1)+1}
=:\frac{Q(\alpha)}{S(\alpha)}.
$$
Since both numerator and denominator have negative slopes and
$Q(\alpha)<S(\alpha)$ for all $1\leq\alpha\leq\min\bra{s+1,t+1}$, $s\geq 1$
and $t\geq1$, then the maximum possible value is achieved when $\alpha=1$.
The lower bound is obtained by evaluating $Q(1)/S(1)$ 
and substituting $s:=r$ and $t:=a-1$.

{\em Upper Bound:}
Similarly, the maximum possible rank is achieved when $\brank{N}=\vartheta$.
Substituting in (\ref{e:R}) $\brank{N}$ by $\vartheta$, and $n$ by $|\mathcal
P|$ we get:
\begin{align*}
R \leq& \frac{\alpha^2+\alpha\inb{st-s-t-1}-ts^2~~~~~~~~~~~~}
{\alpha^2+\alpha\inb{st-s-t-1}-ts^2-t^s-ts}=\\
=& 1+ \frac{t^2+ts}{\alpha^2+\alpha\inb{st-s-t-1}-ts^2-t^s-ts}.
\end{align*}
Note that we can write the denominator as a polynomial function
$Q(\alpha)=\alpha^2+\alpha(st-s-t-1)-ts^2-t^2-ts$, which has minimum value at
$\alpha_\text{min}=-\frac{1}{2}(st-s-t-1)$, which satisfies
$\alpha_\text{min}<1$. Then, since the valid values of $\alpha$ are
$\alpha\in[\min\{s+1,t+1\}]$ and the function $Q(\alpha)$ is monotonically
increasing in this interval, the maximum rate $R$ is obtained when
$\alpha=1$.  The upper bound is obtained by evaluating $R\leq
1+(t^2+ts)/Q(1)$ and substituting $s:=r$ and $t:=a-1$.
\end{IEEEproof}

\vspace{-4mm}
\begin{remark}
The maximum rate of an ($r,a$) $pg$-BLRC code is achieved when the partial
geometry $pg(s,t,\alpha)$ is a generalized quadrangle ($\alpha=1$).
\end{remark}

\begin{figure}
\centering
\includegraphics{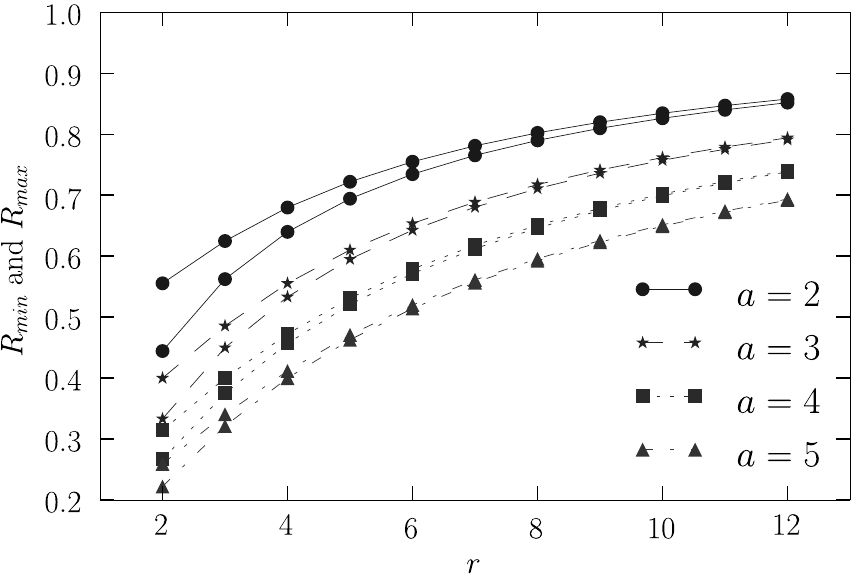}
\caption{\em Upper bound $R_\text{max}$ and lower bound $R_\text{min}$ for the rate $R$ achieved by $pg$-BLRC codes with
repair degree $r$ and repair alternativity $a$.}
\label{f:teo}
\vspace{-6mm}
\end{figure}

In Figure~\ref{f:teo} we use the bounds from Theorem~\ref{t:R} to depict the
minimum and maximum\footnote{Limited to the cases where $r+a-1\equiv1\mod2$.}
possible theoretical rates of an $(r,a)$ $pg$-BLRC code with $\alpha=1$ for
different combinations of $r$ and $a$ values. In general it is interesting to
see how the rate decreases when we either (i) decrease the repair degree $r$,
or (ii) increase the repair alternativity $a$. It means that the two main
objectives to achieve efficient repair mechanisms (small $r$ values and large
$a$ values) entail an increase in the storage overhead of the system (low
rate $R$), posing an optimization trade-off for storage system designers.

\subsection{Explicit $pg$-BLRC Code Constructions}

Note that there is no known construction of generalized quadrangles for all
possible $s$ and $t$ values, and until now, only a few generalized
quadrangles are known~\cite{colbourn2006handbook}.  They are those with
$(s,t)\in\{(2,2), (2,4), (3,3), (3,9),$ $(3,5), (4,4),$ $(4,6), (4,8),$
$(4,16)\}$, and those with $(s,t)\in\{(1,z),$ $(q-1,q+1),$
$(q,q),$ $(q,q^2),$ $(q^2,q^3)\}$, for integers $z$ and prime powers $q$, and
their dual constructions. If we evaluate all the possible $(s,t)$ pairs and
filter out the cases where $R\leq 1/3$ or $n>100$, which are the rates and
length interesting from a practical point of view, then we have 
possible BLRC codes for the following pairs of parameters:
$(r, a)\in\{(r,2),$ $(2, 3),$ $(4, 3),$ $(3, 4),$ $(5, 4)$, $(4, 5)\}$.

\section{Conclusions}

In this paper, we presented a new approach to design locally repairable
codes. Instead of only focusing on codes achieving minimum repair locality and
maximum rate, we analyze how to increase the diversity of this repair
locality, by providing  more than one local repair alternative for data
blocks that need repair. We present an explicit construction of locally
repairable codes that provide different trade-offs between repair locality
and number of repair alternatives per failure. We also provide an upper and
lower bound on the attainable rate of such codes. 

\section*{Acknowledgments}

The authors would like to thank C. Bracken for his valuable
comments. The research of L. Pamies-Juarez, H.D.L. Hollmann and
F. Oggier is supported by the Singapore National Research Foundation
under Research Grant NRF-CRP2-2007-03.

\bibliographystyle{IEEEtran} 
\bibliography{cites}

\end{document}